# Comprehensive Mapping of Continuous/Switching Circuits in CCM and DCM to Machine Learning Domain using Homogeneous Graph Neural Networks

Ahmed K. Khamis
EE Dept., AASTMT, Alex, Egypt
ECE Dept., University at Albany SUNY
Albany NY, USA

Mohammed Agamy
ECE Dept., University at Albany, SUNY
Albany NY, USA

*Abstract*: This paper proposes a method of transferring physical continuous and switching/converter circuits working in continuous conduction mode (CCM) and discontinuous conduction mode (DCM) to graph representation, independent of the connection or the number of circuit components, so that machine learning (ML) algorithms and applications can be easily applied. Such methodology is generalized and is applicable to circuits with any number of switches, components, sources and loads, and can be useful in applications such as artificial intelligence (AI) based circuit design automation, layout optimization, circuit synthesis and performance monitoring and control. The proposed circuit representation and feature extraction methodology is applied to seven types of continuous circuits, ranging from second to fourth order and it is also applied to three of the most common converters (Buck, Boost, and Buck-boost) operating in CCM or DCM. A classifier ML task can easily differentiate between circuit types as well as their mode of operation, showing classification accuracy of 97.37% in continuous circuits and 100% in switching circuits.

## I. Introduction

AI algorithms are used to model computationally complex systems or systems/processes with significant parameter uncertainties. Modern improvements in computation resources enable the incorporation of AI algorithms in power converter design and control. Complex nonlinear problems such as thermal and electromagnetic designs, modeling of layout parasitics and estimation of component stresses under different operating conditions are some areas where AI algorithms can significantly simplify and optimize the design process. [3]–[6]. Power electronics applications of ML have focused on control, component design and maintenance [7]. ML-based surrogate/black box models are used for online prediction tasks to reduce computational effort, memory and power used by classical simulation/mathematical-based models [7]. Design optimization is an additional target as ML models obtain the optimal target without compromising other design constraints or trade-offs of design, which is known in its mathematical formulation as Pareto front [8]. ML-based circuit design should be able to reflect circuit component connectivity as well as the effect of varying the values of these components. In [9] a graph representation of circuits with a combined feature map for input and output nodes was proposed. However, it does not represent details of component types or connectivity, rather it is just a numerical input/output transfer characteristic of the circuit. Reinforcement Learning (RL) was introduced in [10] to optimize passive component values. An updated version of the RL agent was presented in [11], where the RL-based optimization algorithm is used to optimally size transistors. In this case, based on a given design flow, the RL algorithm updates the node embedding in Graph Neural Network (GNN) representation of the circuit to maximize the cost function. One-hot encoding is used to represent transistors, in addition to other internal parameters, which are passed as features to a Graph Convolution Network (GCN) to extract node embedding. Despite the simplicity of this approach, it it is incorrect and does not guarantee a solution in the inverse problem. In other words, in circuit synthesis/generation problem, there is no guarantee that the circuit synthesis neural network can transform the generated graph to a physically realisable circuit. Existing methods do not provide a systematic way of circuit feature embedding in GNNs. These models have several limitations including scalability of circuit size (number of nodes and/or components), mapping con-nectivity and identifying component types within the circuit. This paper proposed a systematic approach for electric circuit representation to enable use of ML design or performance prediction tools. This method has the benefits of being scalable and topology agnostic. In this paper the following key contributions are proposed:

- A comparative review of different research attempts in mapping circuits to ML domain including circuit representation techniques, feature assignment, intended task and how components and connections are represented.
- Proposing three possible circuit representation techniques, listing the advantages and disadvantages while providing mathematical reasons for technique selection.
- The circuit representation includes different circuit element types and circuit connection types, without indulging the concept with numerical tuning or the empirical hyper-parameters optimization of ML.
- Proposing a unified (applied to all circuit elements) node feature assignment algorithm, irrespective of number of connections present in circuit or circuit order, while combining the feature maps of the nodes to generate the feature map for the whole graph in a GNN.
- Proposing a dataset generation algorithm, that is easily applicable to the ML task or application, capturing circuit performance variables of interest in a standardized data format that can be used in ML problems.
- A proof of concept classifier problem applied to variable structure continuous circuits or switching circuits operating in CCM or DCM is presented. The target ML task covers a wide range of possible tasks or even a combination of tasks including regression, classification and clustering tasks, whether it is supervised or unsupervised tasks.

The proposed mapping approach enables a wide range of possible ML tasks or a combination of tasks including regression, classification, clustering, and synthesis of power electronic converter circuits.

## II. PROBLEM DESCRIPTION

Neural networks can construct model from training data after being processed in order to obtain features to characterize the built model. In the case of electrical circuits, the process does not have an established methodology or criteria. Problems with interfacing electric circuits to ML tools are highlighted in this section, while different solutions are proposed in next section.

### A. Circuit Structure Representation Problem

The main problem faced when circuits are to be fed to a NN is the fixed size input layer, which has a defined dimension, invalidating the scalability requirement. The workaround proposed in [12] pre-processes a matrix consisting of multiple vectors representing circuit components, so that the input to the Convolutional Neural Network (CNN) is of a fixed size. Eventually, this workaround added more computational overhead and increased training time and computational resources. Moreover, from a circuit standpoint, it is an incomplete circuit model because it has no explicit representation of the circuit structure or the dynamic behaviour or circuit elements interactions. In this paper we lay some foundations on how the physical properties of an electric circuit can be mapped to ML space, as follows:

1) Circuit performance is independent of circuit entering order or elements order variation as long as the connection is kept invariant (isomorphic circuits). This makes the circuit representation Permutation Invariant.
2) Circuit connectivity (series or parallel connections) and circuit elements values define the circuit performance.
3) Circuits may have any numbers of elements and has no upper boundary.
4) For circuits of similar input/output response (e.g. dual circuits [13]), circuit type/connection will be the identifying factor in each case.

The realization of the last three definitions necessitates that the ML input layer be independent of the size of the input dataset. Hence, the representation becomes Scalable.

### B. Dataset Expressiveness Problem

Machine learning algorithms gain knowledge by iterative training. Datasets contain standardized/normalized data according to the nature of the ML task. Neither a generalized and confirmed methodology to handle circuit datasets nor a feature extraction/definition algorithm are defined that independently capture the circuit topology and the effects of component variation. More importantly, a clear measure of dataset expressiveness is absent. Given the circuits in Fig. 2, every class has identical component count however, their performance is different and depends on component values, especially at resonance, and the dataset should indicate that difference.

### C. Neural Network Topology Problem

The physical circuit topology and the influence of parameter variation on its output variables must be clearly expressed by the selected NN topology. As an example, same circuit performance, can be obtained by using dual components [13]. In [14] a model of similar purpose employs CNN and takes placement images as its features. Arguably, Graph Neural Networks (GNN) are superior in capturing the netlist topology, which is a graph. Moreover, GNN is more efficient in feature encoding. For instance, the shape of a transistor can be represented by two real numbers (width and height) in GNN while it requires an array of pixels for CNN. The spatial features can be easily embraced in GNN by taking the location coordinates as features, which are motivations to take the GNN approach.

### III. REVIEW OF CIRCUIT REPRESENTATION TECHNIQUES

This section offers all possible solutions to presented problems in section II, and highlights the flow of work and derivations made from initial problem statements and better explains available solutions by offering detailed comparisons between them. There has been a lot of attempts to better represent circuits in ML domain, which are thoroughly explained in this paper. Moreover, the paper will also highlight why solutions offered are insufficient, ungeneralizable and empirical solutions, which either require fixed layout, huge datasets or extensive training and very complex models.

*A. Circuit Representation Methodologies*

The main problem is to properly encode circuit problem into computer interpretable form, which has been addressed by three modelling techniques, i.e graph theory, Y-Matrix and Bond graph [15]–[17]. A brief is given on every modelling technique, with an expanded illustration on the one used in this paper, wile a comparison between the merits and disadvantage of three modelling techniques are listed in Table II.

*1) Graph Theory Representation*

Graph theory is a mathematical tool used to model complex systems in a simplified way. In the field of power electronics and converters, graph theory has proved to be a powerful tool for representing and analyzing the complex network of components and their interactions.There have been numerous studies in the literature which use graph theory to represent power electronics and converters [18]. The use of graph theory to represent power electronics and converters has several advantages. Graphs provide a concise and intuitive way to represent the components and their interactions. Furthermore, graph algorithms can be used to analyze the system and identify system faults. However, the use of graph theory to represent power electronics and converters also has some limitations. Graphs are limited in their ability to represent complex systems with many components, as the number of nodes and edges increases, the graph becomes cluttered and difficult to interpret. In addition, graph matrices are usually very large and computationally intensive, making it difficult to obtain simulation results in real-time. This can result in inaccurate or unsatisfactory results [19]. Furthermore, due to the complex relationship between the different components in the power system, the graph model may not be able to accurately represent the real-world system, leading to incorrect results [20]. Graph theory cannot account for nonlinearity and non-smoothness. Power electronic converters are nonlinear systems and their circuits may contain high-frequency harmonics, which is difficult to capture using graph theory [21].Finally, when using graph theory to model a power electronic converter, the system needs to be linearized, which may neglect certain important nonlinear effects. This can lead to incorrect results and further limitations to the accuracy of the model [20].

*2) Y-Matrix Representation*

The admittance matrix is a powerful tool used to represent power systems and power electronic converters. This method of representation has been used since its inception in the 1960s, and continues to be an efficient and novel way to model electrical systems. The admittance matrix is a complex quantity that describes the relationship between the voltage and the current in an electrical network. It consists of a matrix whose elements are admittances of electrical components such as resistors, capacitors, and inductors [22]. This relationship between the voltage and the current provides a useful representation for solving electrical circuit problems [?]. The admittance matrix has been used for many applications such as transient analysis and stability analysis. In particular, it has been used to study power systems [23]. In power system analysis, the admittance matrix can represent the components in the power system such as transmission lines, transformers, and loads, which can be analyzed in both the frequency domain and the time domain [24]. The advantage of the admittance matrix is that it is computationally efficient and provides a concise representation of the wide range of power system components [25].The admittance matrix has also been used for analyzing the stability of power electronic converter systems [26]. Power electronic converters are devices used to convert AC power to DC power or vice versa, and they generally consist of power switches, capacitors, and inductors [27]. Using the admittance matrix, the stability of the power electronic converter can be accurately analyzed in the frequency and time domains [28]. This method of representation is relatively old but can provides for accurate and efficient simulations of power electronic converters.

In a preliminary attempt of this work, different circuits were modelled utilizing Y-bus admittance matrix, where nodes represented buses and admittances serve as node features, while edges represent whether a connection is established between nodes. Fig. 1 shows a three and four element bus systems and its equivalent Y-bus admittance matrix and the corresponding features. However, this representation was proven to be non expressive based on the fact that it is not uniformly scalable, i.e a three and four element (admittances) systems can both have the same number of nodes, which in this case is two, hence losing a very important feature in graph notation. This is because the branch elements are lumped together into a single equivalent admittance making it impossible to distinguish between different elements. Moreover, with this representation, the change in node feature values doesn't discriminate between whether a new element is added or component value has changed.

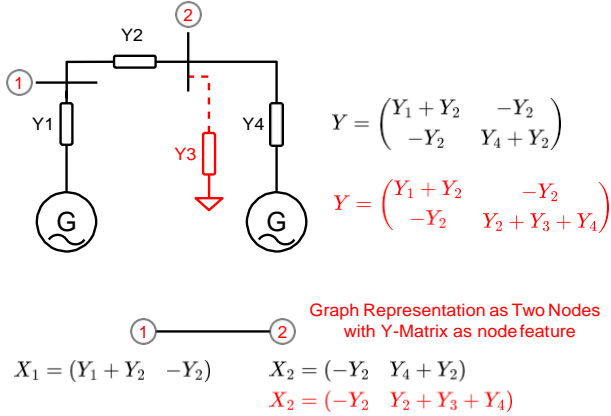

Fig. 1 – Early attempt of converting circuit to graph by using Y-Matrix

*3) Bond Graph Representation*

Bond graphs (BG) were proposed as a graphical language and systematic representation, to overcome limitations of block diagram models [29]. Using BG, a circuit can be modeled as bonds during all possible series and parallel connection permutations and combinations. Two key model elements were devised the 0 junction that is used to represent a parallel connection and 1 junction for series connections [29], [30]. In addition to electric circuits, this approach can be extended to mechanical and chemical models as well [31]–[33]. The BG representation capturing the dynamics of a system is based on transforming (mapping) system components to their BG model counterparts. The bond graph analogies used to describe physical systems in the form of bonds and paths are listed in Table I.

Bond graphs in opposition to transfer function which are behavioral models, belong to the class of structural models. Controllability and structural observability are applicable to BG, which are structural properties of models [37]. Moreover, it was proven in [36] that BGs are structurally identifiable, which allows a unique set of parameters to associate with given input/output response. In other words, bidirectional transformation governs circuit to graph and graph to circuit transformation and hence, graphs generated from ML algorithms can be translated into a circuit if they match structural identifiability criterion.

## IV. REVIEW OF NEURAL NETWORK TOPOLOGIES

### A. Classical Neural Network Topologies

Linear regression, random forest (RF) and artificial neural networks (ANN) are classical regression models used as attempts for regression tasks. For classification tasks, support vector machine (SVM), K-Nearest-Neighbor (KNN) algorithm and RF are used. Convolutional neural network (CNN) and recurrent neural networks are extensively used in ML tasks. CNN models are composed of convolutional layers and other basic blocks such as non-linear activation functions and down-sample pooling functions. While CNN is suitable for feature extraction on grid structure data like 2-D image, RNN is good at processing sequential data such as text or audio [38] due to their ability to leverage statistical properties of the image as euclidean data such as stationarity and compositionality through local statistics. On the contrary, non-Euclidean data has no familiar properties as global parameterization, common system of coordinates, vector space structure, or shift-invariance. Operations like convolution that are taken for granted in the Euclidean case are even not well defined on non-Euclidean domains [39]. From that prospective, it is necessary to use an ML topology that can better represent non-euclidean structures like electric circuits.

### B. Graph Neural Networks

GNNs are composed of definite function layers, but unlike other neural networks, the input is a graph. Acyclic, cyclic, directed, and undirected graphs can be processed by GNN as was stated in the first GNN model in [40]. Scalablity and permutation invariance are unique properties in GNNs allowing input layer to be variable while graph node re-ordering will not affect the NN layer output, which satisfies the requirements needed for physical circuits representations. RNNs and GNNs, capable of directly processing graphs with labeled nodes and edges. An image classification task showed that GNNs outperforms RNNs, both in terms of accuracy and error rate [41]. Convolution operation on graphs is defined by spectral and spatial operations. In [42], spectral-based GCNs was proposed, which used the spectral graph theory to develop a new variant of graph convolutional operation. Graph mutual dependence complexity was solved using non-recursive layers presented in [43]. Moreover, spatial GCNs have been developed based on the fact that spectral GCNs are difficult to extend to large-scale graphs [44]. This makes GNNs suitable for circuit representation.

*1) Graph definition*

Graph G is a defined as (V, E) with V the set of vertices/Nodes equals $v_1, ..., v_N$, while set of Edges E $\subseteq V \times V$. Let N and M be the number of vertices and edges, respectively. Each graph can be represented by an adjacency matrix A of size N $\times$ N : $A_{i,j} = 1$ if there is an edge from vertex $v_i$ to vertex $v_j$, and $A_{i,j} = 0$ otherwise. Every edge has a set of edge features $e$

Table I – Bond Graph terminologies [34]

| Terminology | Description |
|---|---|
| Strong Bond | A single bond that causes effort in the 0 junction and flow in the 1 junction Passive Element A one port element that stores input power as potential energy (C-element), as kinetic energy (I-element) or transforms it into dissipative power (R-element). |
| Causal BG | A BG is called causally completed or causal if the causal stroke known as causality is added on one end of each bond |
| Causal Path | A sequence of bonds with/without a transformer in between having causality at the same end of all bonds or a sequence of bonds with a gyrator in between, and all the bonds of one side of the gyrator having same end causality while all the bonds on the other side with causality on opposite end. That means gyrator switches the direction of efforts/flows on one of its side [9]. A causal path can be a backward or forward or both depending upon the junction structure, elements and causality |
| Branch | A branch is a series of junctions having parent-child relationship. Two differ-ent sequences of junctions can be connected with a common bond or two-port element. Thus, one of the junction's sequence acts as parent branch and the other one as child. |
| Causal Loop | A causal loop is a closed causal path with bonds (of the child branch) either connected to a similar junction or two different junctions of the parent branch |

Table II – Comparison between different circuit representation techniques

| Method | Representation Methods | Merits | Drawbacks |
|---|---|---|---|
| Graph Theory | Component terminals are nodes. Circuit Elements are edges. | Multi-discipline physics based modelling technique. More intuitive graph for human reader. | Converter modelling foundations (duty cycle, CCM & DCM ..etc) are missing/never been addressed No research on graph identifiability from graph to circuit. Circuit graph can be defined using three matrices as shown in [35]. |
| Bond Graph | Elements and connections are nodes with different attributes | Solid foundations on circuits/converter modelling in CCM & DCM. BG is a linear transformation and is mathematically identifiable as shown in [36]. Multi-discipline physics based modelling technique. Generated graph can be defined with one Adjacency matrix. Maintains causality invariance of the system for any operational mode, i.e the state vector resulting from state equation of the system does not change for any operating mode. | Non-intuitive modelling technique. Added complexity of causality assignment. Can yield a bigger graph than graph theory method. |
| Y admittance matrix | Circuit buses are nodes. Connections between buses are edges | Well known methodology for circuit representation. Number of circuit sources can't be extracted. System components can be lumped altogether and information about element count is lost. | Used only for power system representation. Node count is independent from number of components. |

## V. REVIEW OF CIRCUIT REPRESENTATION AND DESIGN USING GNN

In [45] it was shown that the most intuitive way to represent circuit, netlists or layouts is graph representation. It was also stated that graph neural networks (GNNs) are an opportunity replace shallow methods or mathematical optimization techniques, and Table III shows the state of the art circuit representation trials. Many research has utilized GNN in circuit optimizations/classification operations and in many applications like transistor sizing, capacitor value optimization and many more. In [46], [47], the model leverages reinforcement learning (RL) to learn the optimal policy for best parameter selection by rewarding the model for the best Figure of Merits (FOM) composed of several performance metrics. The circuit is embedded into a graph whose vertices are components and edges are wires, while generating a vector for each transistor and passing the graph to the RL agent. Finally, the RL agent processes each vertex in the graph and generates an action vector for each node, then process the graph with an action vector with the purpose of maximizing the reward. [48] proposes a model that solves the forward and inverse problems. In which, the model maps a given circuit to the corresponding transfer function and vise versa. Inversely, the model utilizes gradient descent to optimize the circuit parameters to produce a transfer function. The model leverages the differentiable nature of the neural network and applying gradient descent methods to optimize the input parameters of the neural network. However, the neural network is trained for a particular circuit topology, and hence cannot be used for general circuit representation, in addition to the lack of switching circuit representation. Moreover, [49] proposed a technique for combining the feature maps of the nodes to generate the feature map for the whole graph in a GNN. By propagating information from nodes to nodes representing input and output instead of pooling operation. The paper represents graphs as a concatenations of the feature maps of the input and output nodes. In resonator circuits applications, [49] introduced a model that learns to simulate electromagnetic properties of distributed circuits. Circuit were mapped on system level basis, such that each node refers to a resonator and each edge refers to the interaction between a pair of resonators (i.e., the electromagnetic coupling) between a pair of resonators. This representation does not incorporate the resonator internal structure or if the system had different resonators with different characteristics. By propagating information from nodes to nodes, while representing circuits as concatenation of input and output node features instead of pooling operation, regression task is utilized to obtain predictions about circuit performance. On the other hand, feature concatenation is not the correct technique to represent circuit. Feature concatenation is a numerical representation of circuit inputs and outputs that properly tuned by minimizing the loss function. Attempts has been made to include different circuit topologies and obtain predictions as in [50], where two circuit types were included in the study: the ladder circuits and two stage operational amplifier circuits, with 20k training data instances of resistor ladders with 2 to 10 branches with equal distribution weight. The model is based on DeepGEN architecture and was able to make predictions on ladder circuits with higher number of branches. However, the model's ability to generalize and applicability to other circuit topologies and types remain questionable. Moreover, no clue was given on how to distinguish connection type, and its effect on circuit performance. Moreover, the representation was limited to transistors, without the inclusion of other circuit parameters or elements(Transistor/resistor/voltage sources, .. etc). Also, no guidelines/rules were given on how to model circuit elemtents properties like frequency, phase shift, .. etc. One major drawback in this representation is the elements with multiple terminals like transistors are represented as four connected nodes, which can cause unnecessary excessive computations . In [51], heterogeneous GNN were utilized to construct a graph based on a circuit schematic, where each device (transistor, resistor, capacitor, etc.) can be mapped into different node and edge type within the graph. The model target is to predict net capacitance, which was achieved by mapping connections as nodes with corresponding node information (i.e. net capacitance), preventing information loss if nets were represented as graph edges. To complete the structure, circuits were represented as multi-graphs, where graphs have two edges with opposing directions, and are mapped between every net node and the appropriate device nodes corresponding to terminal connections within the schematic. Despite leveraging heterogeneous GNNs to differentiate between circuit elements nodes and netlist nodes, this representation works around the circuit connection type problem (series or parallel) in the netlist nodes by assigning four types of connection signal (Net to transistor gate, transistor gate to net, Net to transistor drain, and transistor drain to net), resulting in an over complicated representation that extensively require more time at training. Physically, connections in series share the same current and connections in parallel share the same voltage, which are not shown in multi-graph heterogeneous graphs. In the area of analog circuit layout automation, [52] showed a GNN based model that can identify symmetry constraints in analog circuits That can be extended to other pairwise constraints. However, the graph representation of circuits is simplistic as it treats device instances and device pins as graph nodes, while edges represents connections between pin nodes of devices. Eventually, this simplistic representation creates a problem of isomorphic graphs, which was mitigated by adding an additional a two-dimensional vector to node feature to distinguish between whether a node is a device or a pin, which eventually increases computational cost at training. Followed by [53] in which circuits was represented as heterogeneous multi-graphs to the purpose of modelling active and passive elements for analog and mixed signal circuits. In this representation, four types of edges (To transistor (drain), To transistor (source), To transistor (gate), To passive device) are used to represent connections between device/circuit elements, which were represented as nodes. Circuit representation in previous research can be summarized as:

- All methods for circuit to graph representation are arbitrary, without any mathematical/scientific base.
- These methods disregards mapping the connection type and hence is substituted by a significant increase in the number of hidden layers, number of neurons, training for many epochs, ... etc.
- Other implications of disregarding connection type in previous methods are the limited scope of the methodology. Previous methods cannot be applied to any circuit except what it is intended for.
- All methodologies had deficiency in modelling

Table III – Review of circuit representation in previous research

| | Node Features | Edge Features | Circuit Representation | | Task | Network type |
|---|---|---|---|---|---|---|
| | | | Circuit components | Connections (Series/Parallel) | | |
| [46] | DC operating points, One-hot encoding of simulation step, Transistor parameters, Internal capacitances | Featureless | Every circuit element is represented as node, where node features define the element type and DC operating conditions. | | Learning design policy for selecting optimal circuit parameters. | RNN+RL |
| [47] | One-hot encoding of element type Circuit order, Passive and active characteristics | Featureless | No indication was given on connection representation, or its effect on analog circuit performance. | | | GCN+RL |
| [54] | Gate logic level, Controllability, Observability | Featureless | Limited circuit representation in the form of connected nodes according to the physical connection. | | Determine whether an observation point should be added on the output port or not | Meta-path + GCN |
| [48] [49] | Subcircuit coordinates, Center position of the Subcircuit, Angular position of the slit. | Position of the two subcircuits, Gap length, shift | System level representation, where every subcircuit is represented as a node, while edges between two nodes represent distance between two subcircuits. | | Electromagnetic outputs prediction based on resonators relative positions | GCN |
| [55] | Operation type Bitwidth. | Signal information | System level representation, where every node represents a microbench operation, while edges represent signals. | | Operation Delay Prediction for FPGA HLS | GraphSAGE |
| [50] | One-hot encoding of terminal type, Device parameters. | Featureless | Edges, but component terminals are represented as nodes | No direct indication of connection | DC output voltage prediction | Deep-GEN |
| [51] | gate poly length, number of fingers, number of fins, number of copies, length of resistor, Capacitors, number of copies, net N | Featureless | Nodes | No direct indication of connection | Net parasitics Predictions based on physical devices parameters | GraphSage, Relation GCN and Graph Attention Networks. |
| [52] | One hot encoding (Device/Pin) Path based feature | Featureless | Nodes represent component terminals and pins. Components can have multiple nodes representing Pins. Pin/Components are distinguished by node features. Power/GND are represented as I/O nodes. | No direct indication of connection | Binary Classification of layout symmetry | GCN |
| [53] | Node type, Geometry, layer | Featureless | Devices and circuit elements | No direct indication of connection | Binary Classification of layout symmetry | Gated Recurrent Unit based GNN |
| [56] | Device type, Functional Module, Current mirror, Differential pair, Active load, Device dimension, Device location. | Horizontal and vertical distance between pins Pin metal layer, Pin length, Pin type | Nodes with different types | No direct indication of connection | Prediction of IC placement impact on circuit performance | GAT + Pooling (PEA) |
| Proposed | Element ID, Normalized Component Values. | One for continuous Circuits, Duty Cycle for switching circuits. | Nodes with different types | one and zero nodes for every branch/voltage node | Different circuit topologies based ML tasks (Classifier, Regression, Clustering). | GCN + Pooling |

common circuit properties like frequency, phase shift, ... etc.
- Most methodologies mention only elements of interest (Transistors and capacitors), but ignores other circuit parameters like inductance, resistance, voltage source, current source, transformers, ... etc.
- Some methodologies try to simulate the connection type by adding component terminals as nodes and define the circuit as a multi-graph heterogeneous graph. Despite the added complexity and extensive computational cost of heterogeneous graphs, This representation suffers a major disadvantage as different circuit topologies can have the same graph representations (isomorphic graphs). This problem is usually addressed by defining another node feature the define whether a node is a pin or a device at the expense of added computational cost.
- Some representations omits voltage and current

Table IV – Circuit to bondgraph equivalent elements

| Circuit Element | Bondgraph Equivalent Element |
|---|---|
| Voltage Source (V) | Effort Source (Se) |
| Current Source (I) | Flow Source (Sf) |
| Resistance (R) | Resistance (R) |
| Inductance (L) | Inertance (I) |
| capacitance (C) | Compliance (C) |

sources nodes to focus on circuit structure. However, this is incorrect representation since source location can change the circuit behavior.
- Some methodologies include one-hot encoding of device position in circuit along with device type, which inherently means the node features vector size per node is linearly proportional to the circuit size.

## VI. PROPOSED CONVERTER CIRCUITS MODELING FOR MACHINE LEARNING APPLICATIONS

In this section, the proposed formulation of a graph representation of continuous or switching circuits that allow the application of ML algorithms to circuit design and control will be presented. This formulation is completed in several steps:
1) Bond graph modeling of circuit topology.
2) Generating standardized datasets that capture circuit topology, input and output circuit variables and operating conditions.
3) Defining a scalable and permutation invariant NN structure.

### A. Graph Creation Using Bond Graph Modeling

This section explains how to model electric circuit as a graph for further processing.

*1) Continuous circuit presentation as Bond Graph*

An electrical circuit consists of five main components such as resistors, inductors, capacitors, voltage source, and current source. The generalized BG elements and their mathematical relations can describe any continuous circuit and perform analysis of dynamics of electrical systems. Zero-junction is assigned for each distinct voltage node in the circuit where according to Kirchhoff's voltage law (KVL)—the algebraic sum of all voltage drops around a closed circuit is equal to zero. Additionally, one-junction is assigned for each element in the circuit, according to Kirchhoff's current law (KCL)—the algebraic sum of all electrical currents entering and leaving a node is equal to zero), taking into consideration the relative voltage or drops related to each element located between two 0-junctions, since 1-junction represents and effort summation point. Fig. 2 shows the bond graph models of seven classes of resonant circuits of increasing order and Table IV shows the equivalent notations used in BGs with their circuit counterparts.

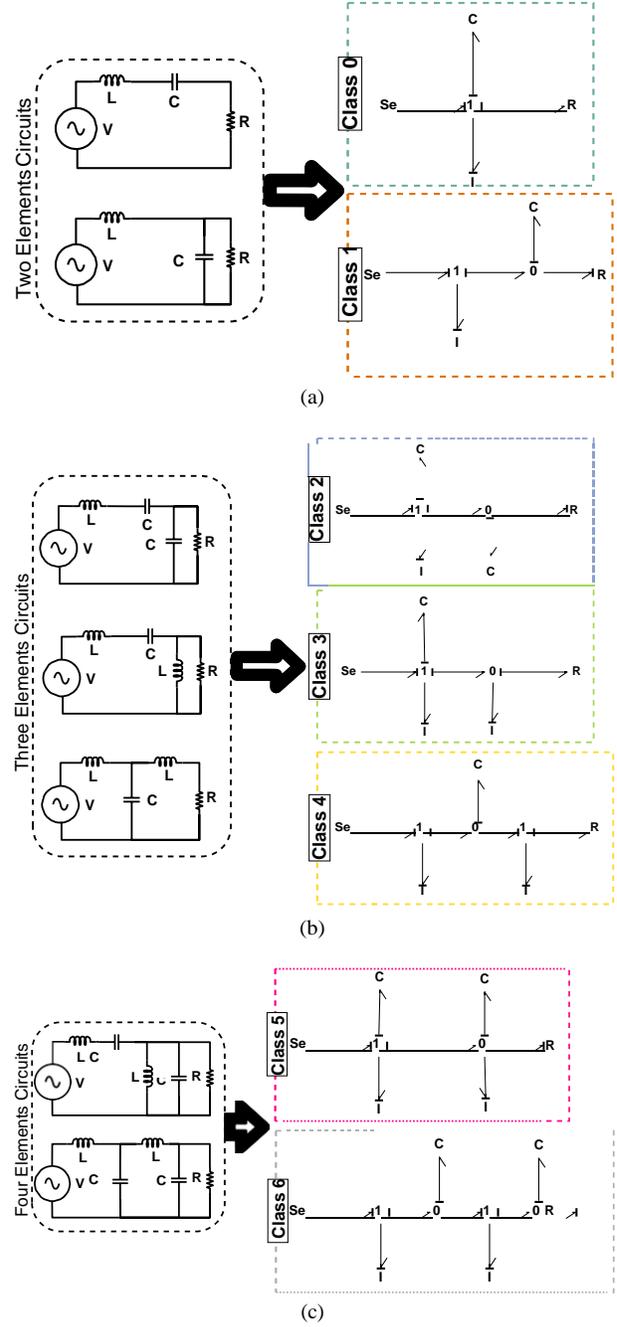

Fig. 2 – Converter circuits to Bondgraphs: (a) Two elements circuits, (b) Three elements circuits, (c) Four elements circuits

*2) Switching Circuit Representation as Bond Graph*

A study in [57], [58] showed that switches (unidirectional or bidirectional) can be represented in BG by the concept of Switched Power Junctions (SPJ) and activated bonds and hence, BG can be used to model switching circuits. Other switch modelling techniques including Modulated Transformer (MTF) with Boolean modulation index m and a resistive element R or the

Ideal Switch Element method where switch state depends on the junction to which the switch element is connected, an energetic connection is established or broken [59], [60]. A comparative study in [61] shows that the most convenient method is the SPJ Modelling method as it does not lead to causality conflicts and leads to a unified model, like the Modulated Transformer method, but does not require additional elements (R) to eliminate algebraic loops. In this paper, the SPJ method will be used to represent switches. Converter topology and its function are defined by the location of the energy storage/resonance elements (L & C) and the type and order of the switching cell. Simplification of Single Pole Double Throw switching cell can be in the form of two Single Pole Single Throw (SPST). Every SPST is modelled as a 1s-junction with two flow decider bonds. For the sake of completion, the physical interpretation of current interruption when the SPST switch is OFF is represented when one flow decider bond is modelled as the zero current source (Sf) and the other flow decider bond is connected to the system. The current source has a zero value, indicating that current falls to zero when switch is OFF. $D$ and $\overline{D}$ are the control signals that control the junction flows. This is uniformly analogous to the duty cycle (D) physical concept in converter circuits. Based on [57], [58], SPST switches combinations can be modelled using (0s and 1s) junctions. Fig. 3 shows a switching cell represented as two SPST switches and its equivalent bond graph representation, the flow decider bond and the zero value flow sources. Additionally, switched power junctions are a generalisation of the already existing zero and one-junction concepts of the bond graph element set [57]. Thus, the traditional zero and one-junctions are special cases of the more general switched power zero and switched power one-junctions. When converters operates in DCM, the inductor current reaches zero before switching cycle is over. This paper utilizes the virtual switch concept to represent converter operation in DCM mode. As the inductor current reaches zero, both switches $S_1$ and $S_2$ are in OFF state. This virtual switch only closes when both switches become OFF. $D_1$, $D_2$, $D_3$ are mutually exclusive control signal to control switches operation. The concept of virtual switch presented in [62] is used to express the converter operation in DCM. This representation is based on the fact that inductor current reaches zero in DCM. The virtual switch shorts the inductor ensuring no current passes through, while connecting certain circuit nodes to maintain voltage balance equations during the DCM time period $D_3$. This representation compatible with the predefined physical property namely Scalability.

### B. Circuits to Graph Representation

The second step is to convert the BG formulation to a graph representation containing all gathered and simulated information including circuit types, classes, nodes, edges, node and edge features. Fig. 4 shows a continuous circuit represented as graph following BG formulation, with minor changes in Switching circuits. Nodes are used to represent circuit element as well as zero and one junctions. Edges are used to describe circuit connection between nodes. Node and edge features describe operating condition of the circuit. In continuous circuits, edge features are set as one describing 100% connection between designated nodes. The same notation is used for switching circuit. Node features are used to describe element type as well as the element value placed in circuit. Some switching circuit properties require special consideration and explained as:

*a) Duty Cycle Representation*

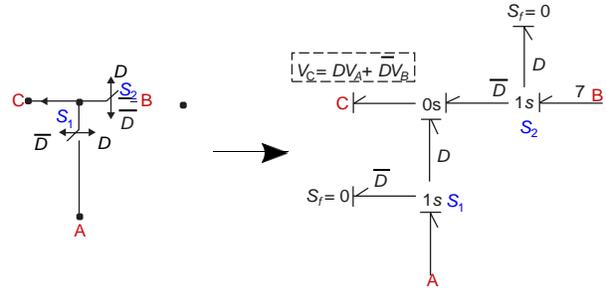

Fig. 3 – Switching cell and equivalent BG formulation

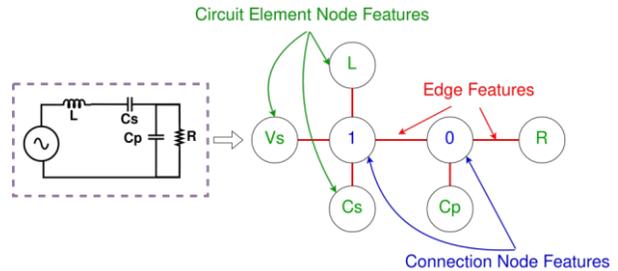

Fig. 4 – Circuit with equivalent BG formulation

The duty cycle is a property in every switching circuit and physically represent the percentage of the connection existence within switching cycle. Duty cycle is mapped as a feature of the edges the connects to switching nodes (0s & 1s nodes).

*b) Switching Frequency Representation*

The one/zero switching junctions representing switching cell are connected to zero-valued current source, interrupting the switch current with frequency equal to switching frequency. In other words, the zero-valued current source works as a control source for every switch. Based on the physical properties of the control source, including the switching frequency as a property of the BG control source aligns with the physical properties of the circuit.

### c) Switching Pattern Representation

A generalized switching pattern representation is proposed, allowing all types of switching patterns and duty cycle variations. This adds more flexibility to represent converters that operate differently when subjected to different switching patterns, i.e resonant converters operating with different control modes. The switching pattern representation is expressed in the control source (flow source in BG representation) node features. Fig. 6 shows two cases of switching patterns. In the first case, the switching is aligned so that the first switching operation compliments the second one. The current source node features should indicate the same phase shift reference, and by default is set to zero. In the second case, where switch operations are not aligned either at turn on or turn off, a phase shift $\varphi$ indicates that delay, and is set the control source of the delayed switch. Combining the phase shift information along with duty cycle information, allows complete representation of the switching patterns in switch operations. Table V summarizes the switching pattern modes and their node feature representation.

Table V – General representation of all possible switching patterns as node features

|        | Representation |
|--------|----------------|
| Case 1 | *Phase shift is set to $\varphi=0$<br>*Edge Features represents duty cycle<br>*Switches which are controlled dependently are represented with the same phase shift. |
| Case 2 | *$\varphi$ Is the phase shift<br>*Delayed switch include phase shift as node feature |

## C. Dataset Generation

Generating a dataset of different circuit topologies, circuit elements and circuit order is shown in this section. Also, a proposed technique for storing recorded data in a general format for any ML task is highlighted. Fig. 7 shows a paradigm for such dataset generation step, where a circuit netlist is converted to its equivalent bond graph model. Since BG is a graph notation for modeling circuits, they inherently have all graph characteristics, with all requirements of graph definitions like number of nodes, node types, edge weights and the adjacency matrix. Finally, BGs are passed to feature assignment algorithm, where features are assigned to each node in graph.

### 1) Feature Assignment

Node features are defined based on circuit element type and its behavior in circuit using the proposed algorithm. Circuit simulations are used to obtaining features describing circuit performance such as node voltages and loop currents. Simulations run for multiple instances at multiple operating points for all circuits including different component values and circuit conditions. Output

Table VI – Feature matrix assignment

**Concatenated Feature Matrix**

| Circuit Element | Element ID |   |   |   |   |   |   | Normalized Value |
|-----------------|---|---|---|---|---|---|---|------------------|
|                 | V | I | L | R | C | 1 | 0 |                  |
| V →             | 1 | 0 | 0 | 0 | 0 | 0 | 0 | xx               |
| I →             | 0 | 1 | 0 | 0 | 0 | 0 | 0 | xx               |
| C →             | 0 | 0 | 0 | 0 | 1 | 0 | 0 | xx               |
| R →             | 0 | 0 | 0 | 1 | 0 | 0 | 0 | xx               |
| L →             | 0 | 0 | 1 | 0 | 0 | 0 | 0 | xx               |
| 0 →             | 0 | 0 | 0 | 0 | 0 | 0 | 1 | xx               |
| 1 →             | 0 | 0 | 0 | 0 | 0 | 1 | 0 | xx               |

values are normalized to common base to avoid sparsity of the feature vector, which is referred in Table VI as "Normalized Values Vector". The proposed feature assignment algorithm is expandable and can include many circuit features if it is desired to be included in the dataset. Therefore, the normalized values vector can be multiple columns listing not only component's value, but also different component properties i.e source frequency in continuous circuits or phase shift in switching circuits. One main function of feature extraction algorithm is to define the circuit element types, which are defines the concept of Element ID. Element ID assigns a binary code based on circuit element type by utilizing one-hot encoding [63]. The second main function of feature assignment algorithm is to concatenate the assigned one-hot encoded vector with normalized values vector, forming the feature matrix of the whole graph with dimension $N \times d_{in}$, where N is number of nodes and $d_{in}$ is the dimension of feature vector.

### 2) Dataset Format

Extracted features and other graph information like types and number of node, adjacency matrix and edge features are saved in a unique graph dataframe format. This unique dataset format features independent graph dataset of circuits, which allows using this graph representation in any ML library independent of saved graph dataset. Since there are many graph ML libraries like pytorch-Geometric [64], DGL [65], Keras [66] .. etc, the final step in the algorithm is to process the dataset to be in a compatible format. Pytorch-Geometric GNN library was chosen to build the GNN structure.

## D. Different Circuit Examples Using Proposed Methodology

This section shows some examples from different areas where the proposed methodology is applicable to many ML applications.

### 1) Example 1: Power System

Power systems (PS) area have a lot of research where ML methodologies has bee applied. Recently GNN has been on the spotlight for application in PS, and many publications utilizing GNN in power systems have emerged. A comprehensive overview of GNN applications such as fault scenario application, time series prediction, power flow calculation, and data generation are reviewed in [67]. In [68], [69] the provided network learns to solve load flow problem on random power grids whose size range from 10 to 110 buses. A method to identify the topology of a PS network is proposed in [70] based on GNN, avoiding errors in Traditional knowledge graphs in the case of errors or informational conflicts in the data. All previously mentioned research empirically transform the PS network into graph without following a circuit-laws-consistent formulation. Fig. 9a shows a PS network example and its graph equivalent with node features, following the proposed methodology.

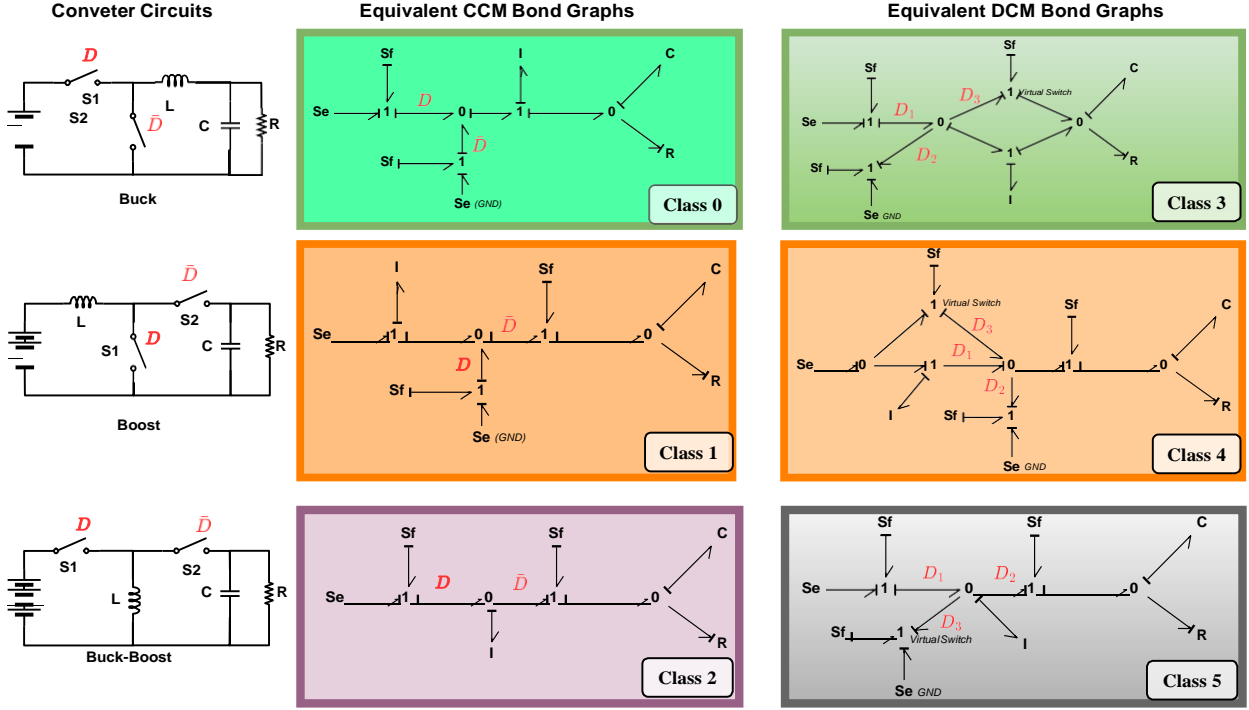

Fig. 5 – Buck, boost and Buck-Boost converters and their equivalent BondGraphs in CCM

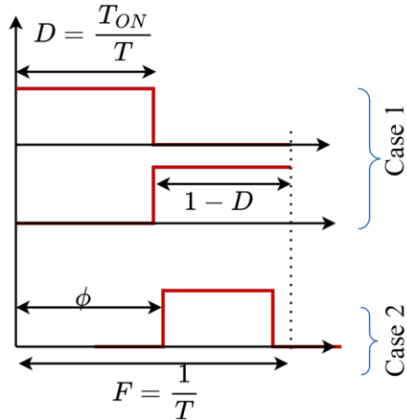

Fig. 6 – Switching pattern representation as features

*2) Example 2: Two-Stage Amplifier*

Fig. 9b shows a two-stage amplifier that was used in [10] as a circuit layout. The equivalent graph representation proposed in this work was arbitrarily transformed into a graph by representing every transistor, resistor and capacitor as nodes connected to each other by edges, disregarding the original connection or the physical/electrical consequences of such connections. The Fig. also shows the proposed graph representation includes component and connection nodes, in addition to node features for each node.

### E. Graph Convolution Network

NN have many variants like GCN [71], GraphSage [72], Gated Convolution [73], Transformer convolution [74] and many more, but the most common is GCN. GCN was chosen for the following reasons:

- Unique ability to extract latent information from graph data compared to other GNN structures as reported in [75].
- Most practical circuit GNN based applications in Table III utilize GCN as their main network model or a part of the model, hence the results from this study can be fairly compared to previous ones.

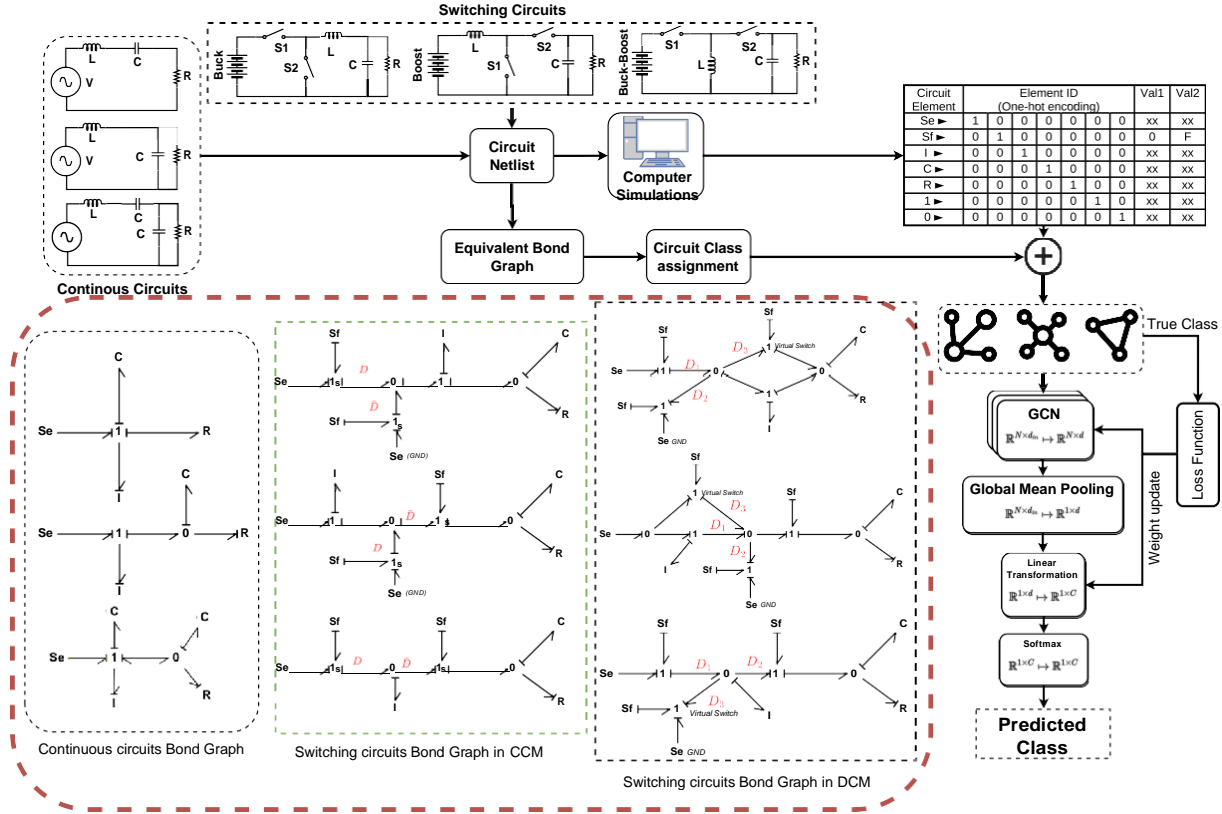

Fig. 7 – From circuit to ML Block diagram

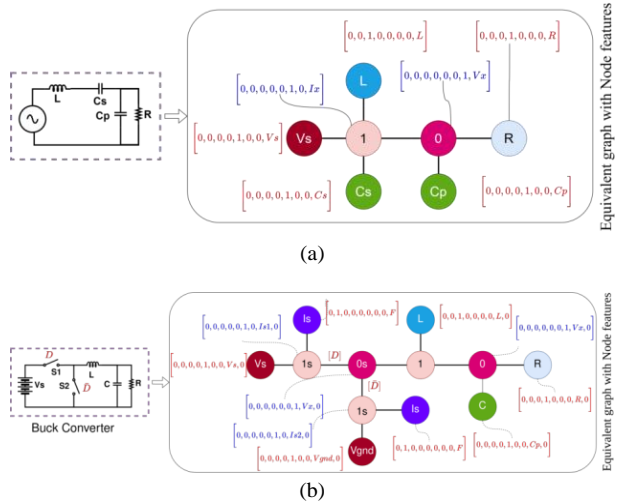

(a)

(b)

Fig. 8 – Equivalent graph with node and edge features in: a) LCC Continuous circuits, b) Buck converter switching circuit

- Simple construction and implementation, which can be beneficial if implemented as digital twin on a microcontroller [76]

The selection of GCN as the engine for the proposed GNN has allowed better focus on other hyperparameters and eventually led to better circuit representation.

GCNs obtain updated features by inspecting neighboring nodes, and aggregating current node information to other neighbours through message-passing process then updating the node state. Eventually, all the nodes in graph obtain knowledge about self and surrounding neighbor information. Fig. 10 shows three layer message passing applied to a single node (node of type 1) of class 1 circuit. A deeper level of neighbor nodes exploration and better awareness of self node position can be gained by adding an additional GCN layer, at the expense of additional computational effort. Three layer GCN network is utilized in this paper as a mid point between exploration depth and computational efficiency. Node features are repetitively aggregated through the GCN layers via multiple message passing layers. At the end of this process, the final node embeddings contain self and all neighbor information.

Mathematically, this initial embedding function is represented by equation (1). The aggregation layer has multiple Graph Convolution Networks (GCN) that performs multiple message passing leaps to collect information about neighbouring nodes and keeps updating the latent dimensional vector with dimension d, which is mathematically represented as in equation (2).

$$X^{(0)} = E(X) \qquad (1)$$

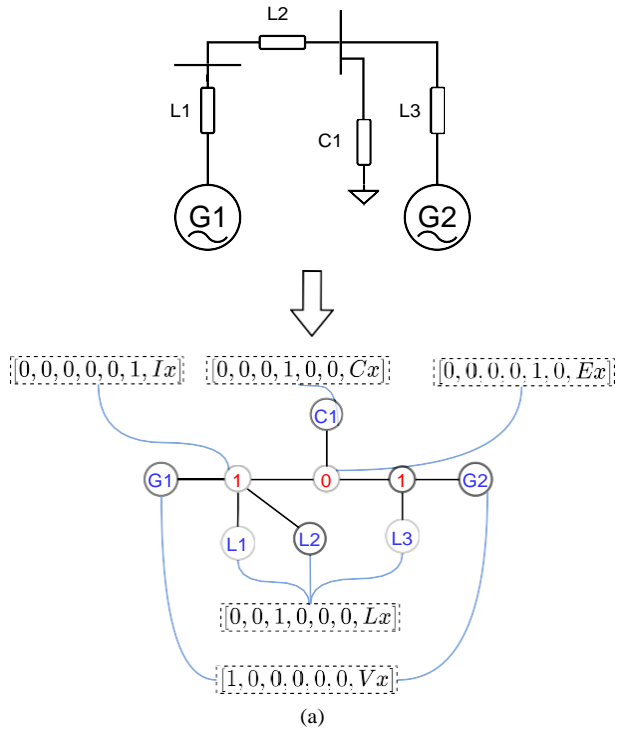

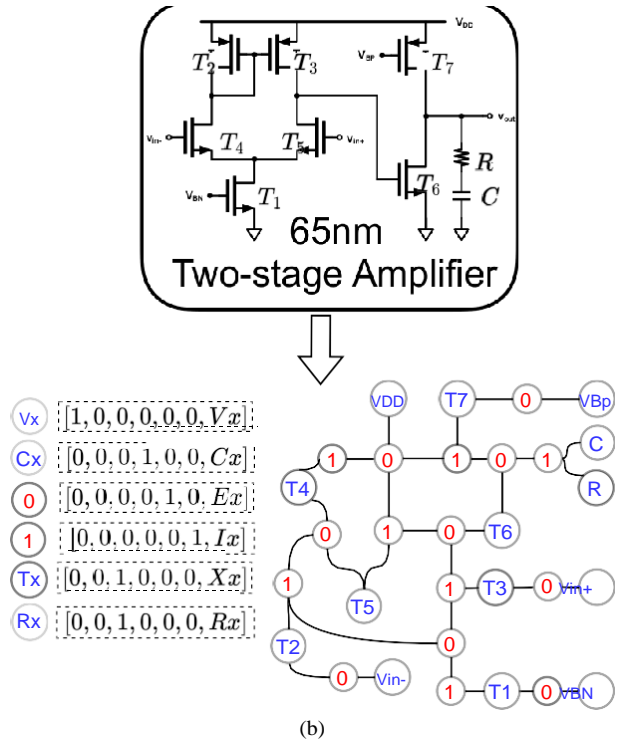

Fig. 9 – Examples of proposed concept in different applications: a) Power system example b) 65 nm 2 stage amplifier example. [10]

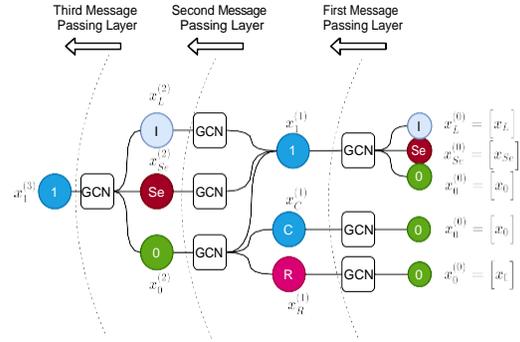

Fig. 10 – Rooted subtree showing message passing applied to node of of type 1 in the circuit of class 1 in Fig. 2 with three GCN layers

$$X^{(l+1)} = \sigma(\hat{D}^{-\frac{1}{2}} \hat{A} \hat{D}^{-\frac{1}{2}} X^l \Theta^l) \quad (2)$$

where $\Theta^l$ is a weight matrix for the l-th neural network layer and $\sigma$ is a non-linear activation function like the ReLU, $\hat{A}$ = A + I, where I is the identity matrix and $\hat{D}$ is the diagonal node degree matrix of $\hat{A}$. This allows the GCN to scale well, because the number of parameters in the model is not tied to the size of the graph.

*F. GCN Time complexity and Graph Scalability Limit*

Generally speaking, there are no limitation on the size of the circuit fed to the ML model (theoretically, the circuit order can be infinite). However, the computation time and RAM consumption are the main concerns when feeding circuit graphs to model, which mainly depends on how the model was built, the libraries used to build the model (pytorch or keras or tensorflow ....etc), the layers depth, operating system used, the model architecture and the output size, ... etc. From a GNN designer prospective, Graph circuit for a GNN input can be represented in two ways:

- sparse: As a list of nodes and a list of edge indices
- dense: As a list of nodes and an adjacency matrix

For any graph G with N vertices of feature vector length F and E edges, the sparse version will operate on the nodes of size N*F and a list of edge indices of size 2*E. The dense representation in contrast will require an adjacency matrix of size N*N, with node degree of d.

The choice of dense or sparse representation not only affects the memory usage, but also the calculation method. Dense and sparse graph tensors require graph convolutions that operate on dense or sparse inputs (or alternatively as seen in some implementations convert between sparse and dense inside the network layer). Sparse graph tensors would operate on sparse convolutions that use sparse operations. Generally, dense computations would be more expensive but faster than sparse, because sparse graphs would require processing of operations in the shape of a list. For simplicity, we assume the node features at every layer are size-$F$. As such, $\Theta^l$ is an $F \times F$ matrix. The time complexity of the convolution operation can be decomposed as:

- Equation (1): which is a dense matrix multiplication between matrices of size $N \times F_l$ and $F_l \times F_{l+1}$. We assume for all $l$, $F_l = F_{l+1} = F$. Therefore, this is $O(NF^2)$.
- Equation (2): which is a multiplication between matrices of size $N \times N$ and $N \times F$, yielding $O(N^2F)$ time complexity. Hence, the neighborhood aggregation for each node therefore requires $O(dF)$ work, with a total of $O(NdF) = O(EF)$. $\sigma(\cdot)$: is the activation function which is an element-wise function, so its cost is $O(N)$.

Over L layers, this results in computational time complexity of:
$$O(LNF^2 + LNdF + LN) = O(LNF^2 + LNdF) = O(LNF^2 + LEF)$$

### G. Optimal Node And Edge Features Exploration

To determine the optimal representation of circuit component values, twelve experiments were performed on the continuous circuits of Fig. 2 and the results are shown in Fig. 11 - Fig. 14. The dataset contained 6000 graphs representing the seven circuit types. 70% of the dataset was used for training. The data is shuffled before being applied to the model, and there was no mutual data between training and testing. Cross entropy loss function is used in training the model with Adam optimizer [77] with learning rate of 0.02. Twelve experiments were conducted in order to obtain conclusions and a paradigm of how the node and edge features should represent the circuit parameters. These experiments were divided into four sets. Each set contains three experiments and a conclusion based on observations from these experiments. The conditions/modifications applied on the dataset when fed to the classifier are listed on the left of each set. The purpose of these experiments is to identify the effect of different component representations, and how would that affect the ML task. Figures also show the classifier problem evolution ranging from three class to seven class classifier problem, along with physical circuit elements representation as features.

The purpose of the upcoming experiments is to explore the highest impact features on task accuracy. However, since features are hyper-parameters, some result obtained from edge features may eventually update how the node features are expressed. In the first set of experiments shown in Fig. 11, edge features are explored and the problem is limited to three classes classifier, edge weights are separately tested as normalized frequency ($\frac{circuitFrequency}{resonanceFrequency}$) vs. being set as ones, vs. being the circuit frequency. This experiment is concluded with the highest accuracy achieved is when edge weights were set as normalized frequency and as ones. As frequency can be included as edge features, it can be tested if capacitive elements can to be expressed as ($\frac{1}{normalizedFrequency}$), which is the purpose of the second experiment set.

Fig. 12 is the second set of experiments, where edge weights were set as the normalized frequency, while nodes that represents capacitive elements were set to have ($\frac{1}{normalizedFrequency}$) as edge feature. Another experiment is to test whether negative component values would increase the accuracy, or setting the capacitive components as $\frac{1}{C}$. These experiments are reflection from circuit analysis as $X_c = \frac{-j}{Frequency \times C}$. However, the results shows that negative capacitive element value and its edge feature as ($\frac{1}{normalizedFrequency}$) have negative effect on the accuracy of the classifier, while setting capacitive elements as of inverted value ($\frac{1}{C}$) had a significant training accuracy boost to 91.12%. It is imperative to modify node features expression for capacitive elements. Eventually, circuit graph dataset was modified to include this change in the the third experiment set. Also, from the first experiment set, edge features set as one had the highest accuracy score. The next experiment aims to explore if the concluded node and edge features modifications can enhance the accuracy.

In the third experiment set, the highest accuracy of 100% was achieved in training and testing when edge weights were set to ones and capacitive elements has node feature values of ($\frac{1}{C}$). The first experiment tested whether edge feature can be used as a scaling factor substituted by the node feature. The second one tested whether edge weights can be set to one, while the third experiment tested if inductive elements can be set as ($\frac{1}{L}$). From results shown in Fig. 13, it can be concluded that utilizing edge features for scaling deteriorates the classification accuracy as well as representing inductive elements as ($\frac{1}{L}$). The optimal edge feature can be defined to be one, without embedding any circuit characteristics or parameters.

In the last set of experiments in Fig. 14, all outcomes and recommendations that was concluded from previous experiments were taken into consideration, while increasing the classification problem difficulty to four, five and seven classes classification problem to further verify the optimal representation. In a four-classes problem, the classifier scored a training accuracy of 92.3%, while in five-classes problem the training accuracy score was 95.92%. Lastly, the seven-classes problem resulted in training accuracy score of 97.37%. The discrepancy of accuracy scores while using the same feature representation is due to the change in dataset number of circuits. The result is a graph of a circuit with connection nodes and element nodes each has its own features. Nodes are connected by edges having edge features of one.

### VII. CASE STUDY

As a proof of concept, the proposed approach is applied to map two types of topologies: i) continuous circuits and ii) switching circuits, to a ML compatible representation. Seven resonant circuit topologies of cir-cuit orders ranging from second to fourth order as shown in Fig. 2, and three switching circuit topologies in CCM and DCM shown in Fig. 3 are fed to a classifier to show the applicability of the proposed methodology to any ML task. Following the sequence illustrated in Fig. 7 and same steps presented in this paper and in [1] and [2], converters are converted to graph form and computer simulations are used to assign normalized node features of the generated graph according to section VI-C1. Steady state simulations are run for multiple instances at multiple operating points for all circuits including different component values and circuit conditions and circuit behavior is recorded and stored. The circuit simulation sampling rate is a measure of the accuracy of the circuit simulations in the continuous circuit classifier case. In this case study, a dataset of 6000 graphs with 6000 steady state simulations have been normalized to a common base. This helps to ensure that each feature vector is consistent and not overly sparse. The normalized values vector is then used to provide a representation of the circuit simulation data that is accurate and reliable. To ensure that the sampling rate is accurate, the graphs are divided into a number of subsets based on circuit class, and each subset is simulated separately. Each of these subsets is tested for accuracy, and any discrepancies are noted and addressed. After all the subsets have been tested and corrected, the overall sampling rate of the circuit simulations can be determined. Once the sampling rate has been determined, the normalized values vector is concatunated with element ID to complement the feature vector. Fig. 15 shows a block diagram of the classifier structure. Three GCN layers are used to get information about 3$^{rd}$ level neighbors. The classifier output layer computes a probability score for the class of each topology.

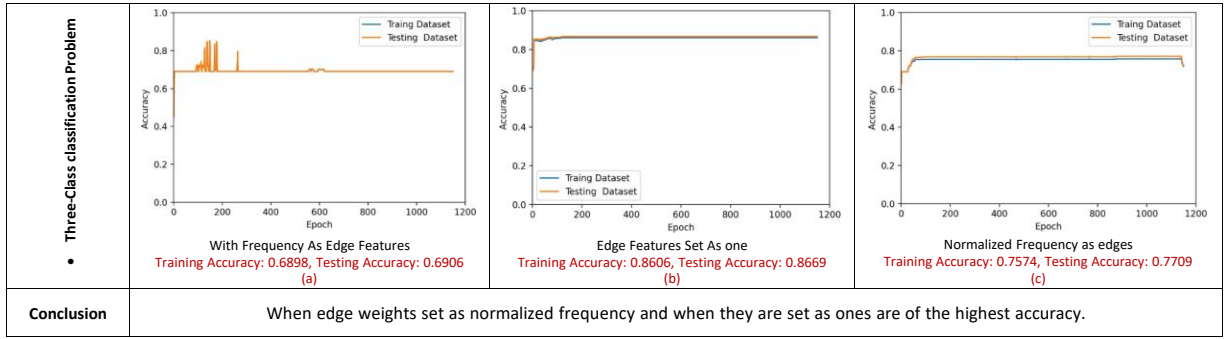

Fig. 11 – First experiment set. Edge weights are set as: a) Frequency, b) value of one, c) Normalized frequency.

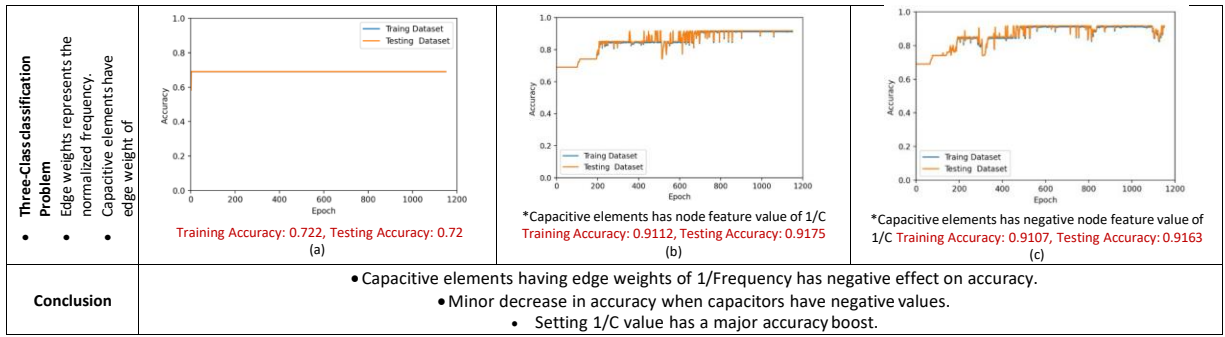

Fig. 12 – Second experiment set. a) No change in node features, b) Capacitive element representation is $\frac{1}{C}$, c) Capacitive element representation is $\frac{-1}{C}$.

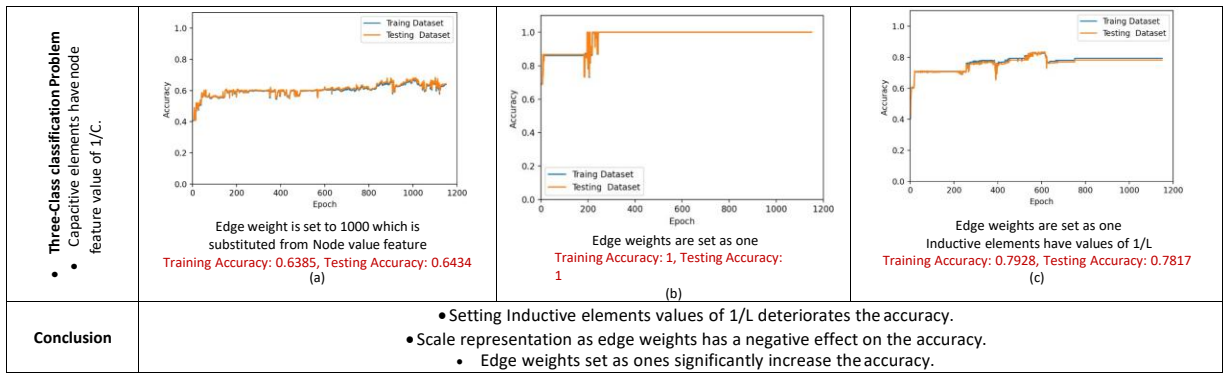

Fig. 13 – Third experiment set. Edge weights are set as: a) Scaling factor, b) value of one, c) value of one but different inductive element representation.

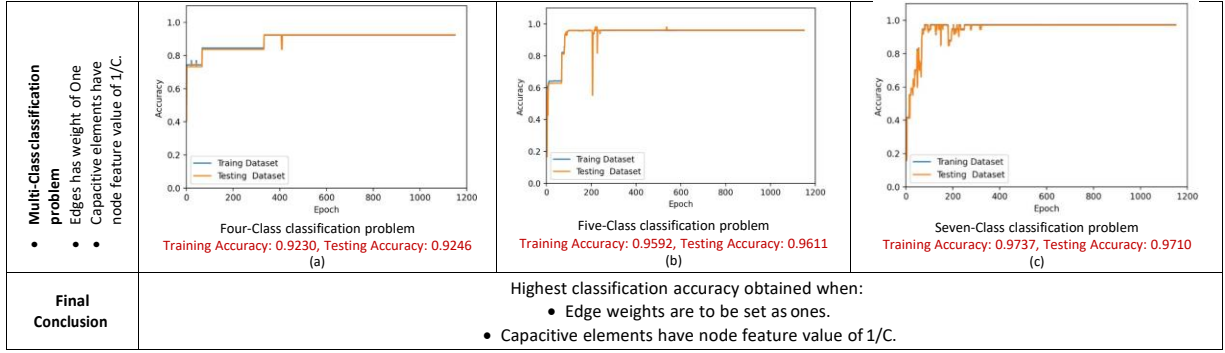

Fig. 14 – Fourth experiment set. a) Four-class, b) Five-class, c) Seven-class classification problem.

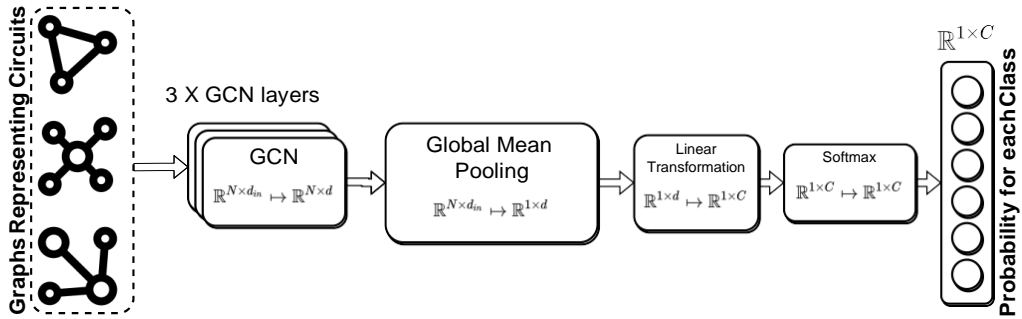

Fig. 15 – Circuit classifier structure [1]

*1) Classifier Problem Formulation*

Circuit topologies in graph forms (G) are fed to the classifier. Each circuit graph has number of nodes (N) along with their corresponding node features (X) each has dimension ($d_{in}$). The adjacency matrix (A) defines connections between each node. The classifier outputs a probability (Y) of a converter to belong to a certain class (C). Sub-GCN networks are embedded in each GCN layer, allowing aggregation processes between feature vectors in the neighboring nodes. Hyperbolic tangent ("tanh") is used as the non-linear activation function, while being slower than the Rectified Linear Unit (ReLU) activation function, it helps to avoid the dying ReLU problem due to the very different values of both inputs and outputs [78]. The global mean readout (GM-Read out) layer returns graph level outputs by averaging GCN processed node features. A fully Connected (FC) linear layer is a score function for each circuit, while (Softmax) output layer is used to calculate the probability, in range of [0-1], of each circuit belonging to a certain class. The Softmax function formula $\sigma()$ is stated in equation (10). The classifier uses training datasets and updates weights or GCN layers and linear layers by minimizing the cross entropy loss function, which is shown in equation (11), where:

- M - Number of classes

- log - The natural log

- Y - Binary indicator (0 or 1) if class label c is the correct classification for observation O.

- p - Predicted probability observation O is of class C.

A mathematical formulation of the transformations of the designed classifier is stated as:

$$Y = classifier(X, A) \qquad (3)$$

Where

$$X \in \mathbb{R}^{(N) \times d_{in}} \quad (4)$$
$$Y \in \mathbb{R}^{C \times 1} \quad (5)$$
$$GCN^{(k)} : \mathbb{R}^{N \times d_{in}} \rightarrow \mathbb{R}^{N \times d}, k \in {0, 1, .., k-1} \quad (6)$$
$$GM - Readout : \mathbb{R}^{N \times d} \rightarrow \mathbb{R}^{1 \times d} \quad (7)$$
$$FC : \mathbb{R}^{1 \times d} \rightarrow \mathbb{R}^{1 \times C} \quad (8)$$
$$Softmax = \mathbb{R}^{1 \times C} \rightarrow \mathbb{R}^{1 \times C} \quad (9)$$

where

$$\sigma(z_i) = \frac{e^{z_i}}{\sum_{j=1}^{K} e^{z_j}} \quad for\ i = 1, 2, \ldots, K \quad (10)$$

$$CrossEntropy = - \sum_{c=1} y_{o,c} \log(p_{o,c}) \quad (11)$$

*2) Results and Analysis*

  *a) Continuous Circuit Classifier*

Training and testing accuracy after 1200 epochs are shown in Fig 16, scoring 97.37% and 97.10 %, respectively. 70% of the dataset containing 6000 graphs representing the seven circuit classes was used for training. Cross entropy loss function is used in training the model with Adam optimizer with learning rate of 0.02. Fig. 17 shows the 2-D embedding of the classifier testing dataset output. It can be clearly seen that graphs falling in the same class cluster together.

The confusion matrix shown in Fig. 18 is used to analyze the classifier behavior and obtain insights about its functionality. The array gives an insight about overlaps/errors in class predictions. Other classifier assessment metrics are listed in Table VII, which shows the precision, recall, F1 and support metrics. The following notations are used to assess binary classifiers performance, but are also extended to multi-classification problems.

- Positive: The graph is classified as a member of the circuit class the classifier is trying to identify.
- Negative: The instance is classified as not being a member of the class we are trying to identify.

True or false can be added to Positive or negative to indicate whether the classifier has correctly predicted the class or misclassified it. Generally, precision is a measure of true positive instances, which shows how many of the positive predictions made are correct. Recall aka sensitivity, is a measure of how many of the positive cases the classifier correctly predicted with respect to the over all the positive cases in the data. The F1 score is the percentage of correct class predictions. A mathematical formulation of the evaluation metrics are listed in equations (13-14).

$$Precision = \frac{TP}{TP + FP}$$

$$= \frac{\text{No. of correct predictions belonging to specific class}}{\text{Total No. of predictions belonging to that class}} \quad (12)$$

$$Recall = \frac{TP}{TP + FN}$$

$$= \frac{\text{No. of correct predictions belonging to specific class}}{\text{Total No. of correct predictions in the dataset}} \quad (13)$$

$$F1score = \frac{2 \times Precision \times Recall}{Precision + Recall} \quad (14)$$

Class two and three have F1 scores of 0.87 and 0.89, respectively. Since F1 score embeds precision and recall into one computation, the weighted average of F1 should be used to compare classifier models, not global accuracy. The Recall of class 2 is 0.77, indicating a misclassification occurs. On the other hand, the Recall score of class three is 1, indicating all class 3 circuits were correctly classified. This analysis indicates misclassification of 52 class 2 circuit graphs as class 3, resulting in a precision measure of 0.8. Additional observations from confusion matrix, classifier metrics and the 2-D vector mapping can be summarized as follows:

- Circuits with similar connections are distinctly classified but the clusters appear close in the 2-D vector mapping. Classes (four and six) are fourth order circuits but are dissimilar in physical connection, hence are mapped in the same vicinity but close. Similarly are classes (Zero and one), follow the same principle. On the other hand, classes (two and three) are second order circuits sharing almost identical circuit connection, hence are mapped very close to each other.
- The same concept is applied to circuits with dissimilar circuit structures, as they are clustered far from each other in the 2-D map i.e classes 0 and 5.
- The similarity between classes two and three in connection and number of nodes causes 2.63% classification inaccuracy. Further tuning of the weights of the linear layer can improve the classifier selectivity.

  *b) Switching Circuit Classifier*

The trained classifier scored 100% for training and testing data, when trained for 200 epochs. In Fig. 20, a 2-D output representation of 1800 test dataset graphs are plotted and colorized according to their predicted class. Circuits of the same topology are distinctly identified and clustered together. Further, the operating mode of each of the circuits (CCM or DCM) is also identified. The different loci of the 2D plot from every class is a result of convolution operation taking all graph properties representing circuits like component values, type and switches duty cycle and converting it to a lower dimension (2-D). It is also noted that graphs of the same converter topology form groups and cluster in

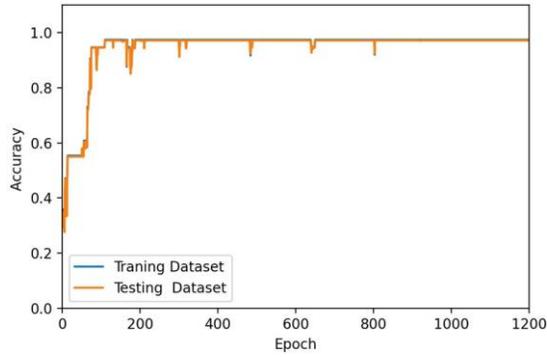

Fig. 16 – Circuit classifier accuracy

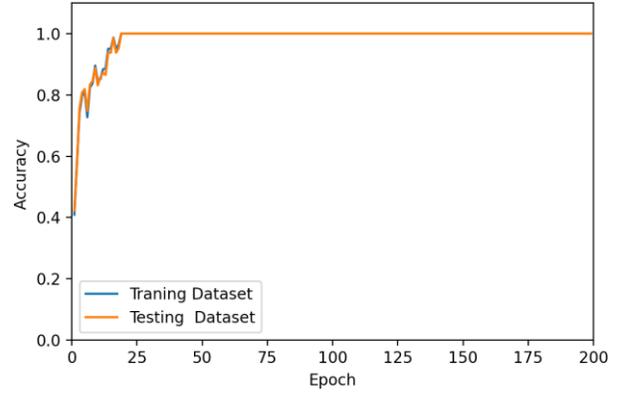

(a)

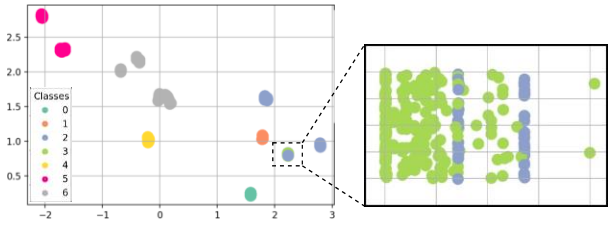

Fig. 17 – 2-D embeddings of circuit graphs

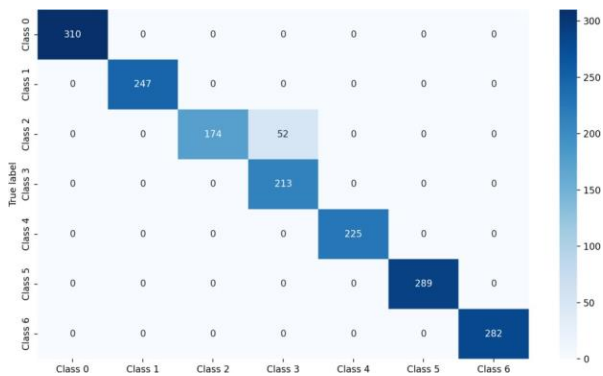

Fig. 18 – Confusion matrix for seven classes circuits

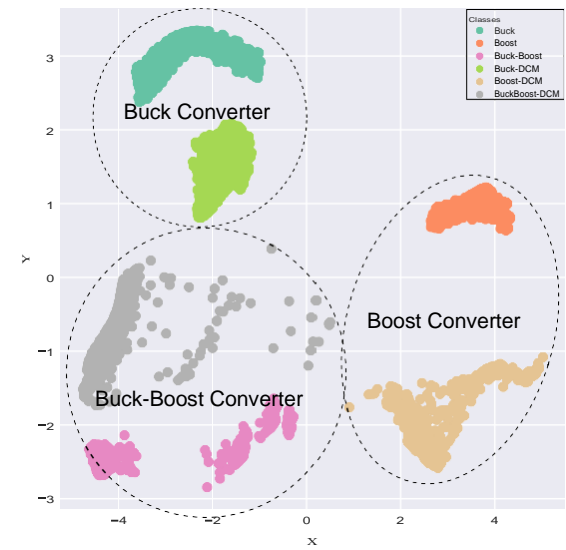

(b)

Fig. 20 – a) training and testing data classification accuracy, b) 2D embedding of the three converters in CCM and DCM after classification

Table VII – Continuous circuits classifier assessment metrics

| Circuit Class | Precesion | Recall | F1 score | Support |
|---|---|---|---|---|
| Class 0 | 1.00 | 1.00 | 1.00 | 310 |
| Class 1 | 1.00 | 1.00 | 1.00 | 247 |
| Class 2 | 1.00 | 0.77 | 0.87 | 226 |
| Class 3 | 0.80 | 1.00 | 0.89 | 213 |
| Class 4 | 1.00 | 1.00 | 1.00 | 225 |
| Class 5 | 1.00 | 1.00 | 1.00 | 289 |
| Class 6 | 1.00 | 1.00 | 1.00 | 282 |
| Macro avg | 0.97 | 0.97 | 0.97 | 1792 |
| Weighted avg | 0.98 | 0.97 | 0.97 | 1792 |
| Accuracy | | | 0.97 | 1792 |

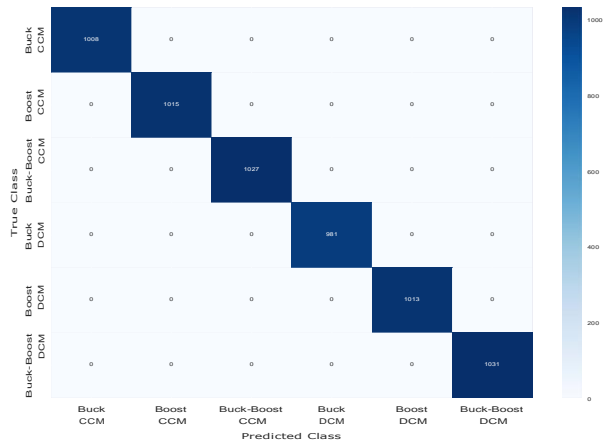

Fig. 19 – Confusion matrix for DC-DC converters

close proximity.

## VIII. DISCUSSION AND FUTURE WORK

This methodology of circuit representation allows incorporating ML techniques in many applications, and can serve the purpose of generating application-specific circuits. Machine learning and neural network models in general are heavily dependent on hyper-parameter tuning. Several aspects are to be included when circuit designer incorporate ML model in circuit design like network depth, number of neuron, activation functions, pooling layers ... etc. These uncertainties in ML models adds more burden when incorporating ML techniques in circuit design. Eventually, a network update is a must at some point of the design process, and eventually designer must fine tweak the ML based design tool. The proposed method can be applied to a wide range of applications such as, power electronic converters condition monitoring and prognostics, since the developed representation maps the circuit structure and thus voltage stresses at each node and current stresses in each branch can be evaluated and tied to a component/converter reliability function. Another application is network structure and fault detection in large power systems [79]. Circuit design is another application that fits the proposed methodology, where circuit performance parameters are set, and the GNN model can generate a circuit topology that meets the input criteria. Moreover, this study can be further developed to for the purpose of linking finite element modelling software in AI assisted design of magnetic components for the purpose of optimal component values/shape design. Additionally, the proposed methodology has very high potential in circuit obfuscation and reverse engineering when it is required to identify/obscure circuit structure [80]. One idea works on the circuit side utilizing the GNN capability of learning the proper transformation function of the converter, i.e can obtain a mathematical transformation of every circuit component and eventually all circuit behavior. On the application side, the end goals whether they are gain, current ripples, magnetic design .. etc, are transformed into a fictitious statistical domain, and the purpose of the GNN is to generate circuits with similar statistical domain. This can be beneficial to train AI to generate application specific converters, which eventually will help reduce component size, increase power density, speed and efficiency. This methodology is also applicable in power system applications such as network reconstruction and fault detection and load flow estimation ...etc.

Table VIII – DC-DC converters classifier assessment metrics

| Circuit Class | Precesion | Recall | F1 score | Support |
|---|---|---|---|---|
| Buck-CCM | 1.00 | 1.00 | 1.00 | 1008 |
| Boost-CCM | 1.00 | 1.00 | 1.00 | 1015 |
| Buck-Boost-CCM | 1.00 | 1.00 | 1.00 | 1027 |
| Buck-DCM | 1.00 | 1.00 | 1.00 | 981 |
| Boost-DCM | 1.00 | 1.00 | 1.00 | 1013 |
| Buck-Boost-DCM | 1.00 | 1.00 | 1.00 | 1031 |
| Macro avg | 1.00 | 1.00 | 1.00 | 6075 |
| Weighted avg | 1.00 | 1.00 | 1.00 | 6075 |
| Accuracy | | | 1.00 | 6075 |

## IX. CONCLUSION

In this paper a graph representation of electric circuits is proposed. This method enables a dynamically scalable interface of different circuit aspects including physical connections, component values and mode of operation, to the machine learning domain. Applying the circuit graphs as inputs to a GNN different circuit modeling, design and optimization tasks can be performed. The effect of bond graph feature selection, scaling and formulation was also analyzed. Optimal feature representation results in a more well defined feature matrix and consequently a more accurate circuit and operating mode identification. As a proof of concept case studies of classifiers of continuous and switching circuits were presented where, the proposed algorithms were proven to distinctly identify with high accuracy circuit types based on physical connectivity as well as identifying their mode of operation based on parameter values and control variable values.